\documentclass[aps,prd,twocolumn,showpacs,groupedaddress,nofootinbib,letterpaper]{revtex4}

\usepackage{amsmath}
\usepackage{amsfonts}
\usepackage{amssymb}
\usepackage{hyperref}

\def\beq{\begin{equation}}
\def\eeq{\end{equation}}
\def\bea{\begin{eqnarray}}
\def\eea{\end{eqnarray}}
\def\ben{\begin{enumerate}}
\def\een{\end{enumerate}}

\def\b{\beta}
\def\g{\gamma}

\def\o{\omega}

\def\del{\partial}

\def\half{{\textstyle{\frac{1}{2}}}}

\def\vphi{\varphi}

\def\C{Ref.~\cite{Carroll2009} }
\def\Cc{Ref.~\cite{Carroll2009}}

\begin{document}

\title{Stability of the aether}
\author{William Donnelly}
\email{wdonnell@umd.edu}
\author{Ted Jacobson}
\email{jacobson@umd.edu}
\affiliation{
Center for Fundamental Physics, Department of Physics \\
University of Maryland, College Park, MD 20742-4111
}

\begin{abstract}
The requirements for stability 
of a Lorentz violating theory are analyzed. 
In particular we conclude that
Einstein-aether theory can be stable 
when its modes have any phase velocity, 
rather than only the speed of light as was 
argued in a recent paper.
\end{abstract}


\maketitle

The purpose of this note is to argue for the
appropriate notion of stability in a theory
with broken Lorentz symmetry 
that supports modes with phase velocities different from the speed of light.
In particular we are motivated by the example of
Einstein-aether theory, but our considerations 
are quite general. More specifically, we 
shall argue that the stability criteria imposed on
this theory in Ref.~\cite{Carroll2009} are 
overly restrictive. The conclusion is that the 
theory is actually stable for an open set in the
four dimensional coupling parameter space rather than
for only a one dimensional subspace.
The issue of stability in Lorentz violating theories
was also addressed in Refs.~\cite{Dubovsky2005,Babichev2007}, 
which include arguments closely related to those advanced here.

Einstein-aether theory is an example of a
theory where Lorentz symmetry is dynamically broken.
Aside from matter, the fundamental fields are 
the spacetime metric $g_{ab}$ and a timelike unit
vector field $u^a$, the ``aether'' 
(see Ref.~\cite{Jacobson2008} for a review).
Flat spacetime with a constant aether is a 
solution to the theory, and linearized perturbations
of this solution satisfy second order hyperbolic equations.
There are modes with five different polarizations: two spin-2, two spin-1,
and a single spin-0 mode. For all these modes,
the frequency $\omega$ 
and spatial wave vector $k$ defined 
relative to the rest frame of the aether
satisfy a gapless dispersion relation, $\omega^2 = v_i^2 k^2$,
where $i$ labels the spin. 
The squared velocities $v_i^2$ 
depend on the coupling parameters
in the Lagrangian, and are generally 
different from each other and different from the
``speed of light" $c$ defined by the 
null cone of the metric $g_{ab}$.
The conditions $v_i^2>0$ impose inequalities
on the coupling parameters, guaranteeing that the frequency
is real, so the perturbations do not
grow exponentially in time if the spatial wave vector is real~\cite{Jacobson2004}.
Another set of inequalities implies that the energy carried by these
modes is positive~\cite{Lim2004,Eling2005,Foster2006}, and these inequalities 
can be satisfied simultaneously with the former stability 
inequalities. (As of yet, no nonlinear extension of this
positive energy result is known, except in the special
case of static spherical symmetry~\cite{Garfinkle2011}.)

It was recently argued in Ref.~\cite{Carroll2009}
that one should 
require the above stability
criteria, i.e.\ real frequency and positive energy, 
not only for modes
with a real wave vector in the aether frame, but more generally
for modes with real wave vector in any Lorentz frame 
defined with respect to the metric $g_{ab}$,
and for energy defined in any Lorentz frame.
In a non-Lorentz invariant theory this is obviously 
a much stronger requirement, and in fact it was concluded
that Einstein-aether theories 
are unstable except for a small number of special cases in
which all modes propagate at exactly the speed of light\footnote{Actually, in \C
the decoupling limit was taken. That is,
the metric is fixed to the Minkowski metric and not
varied in finding the field equations.}.
We shall now argue, however, that these stronger conditions
are not required by stability of the theory, and are not justified
given the structure of the theory. 

In fact the reasoning of Ref.~\cite{Carroll2009} 
applies to any linear theory with modes propagating at different speeds, 
not only to Einstein-aether theory. Also, the 
dynamics of the metric and aether themselves
play no essential role except to define those modes.
Hence we will discuss the simpler 
setting of fields on a spacetime with a fixed Minkowski metric
$\eta_{ab}$ (with signature $({+}{-}{-}{-})$)
and a fixed timelike unit vector $u^a$.
A free scalar field $\vphi$ that propagates
at speed $v$ with respect to the rest frame of $u^a$ is
minimally coupled to the effective (inverse) metric
\beq \label{gv}
g_{(v)}^{ab} = u^au^b+v^2(\eta^{ab} - u^au^b),
\eeq
with Lagrangian density
\beq \label{lagrangian}
{\cal L} = \half \sqrt{-\eta}\, g_{(v)}^{ab}\del_a\vphi\, \del_b\vphi
=\half\Bigl(\del_t^2\vphi -v^2\del_i^2\vphi\Bigr),
\eeq
where the second expression is written in the Minkowski 
coordinate system $(t,x^i)$ of the metric $\eta_{ab}$, 
adapted to the rest frame of $u^a$. We consider this 
model with arbitrary positive values of $v$.

In the context of Einstein-aether
theory, there is good reason to allow $v$ to be greater than $c$.
If the coupling constants are chosen so that the post-Newtonian preferred frame parameters of the theory are in agreement with observational constraints,
then the positivity of the energy (in the aether frame) requires $v \ge c$. 
Also, to satisfy the vacuum Cherenkov constraint for ultra high-energy cosmic
rays, all $v>c$ are allowed, but any $v$ 
less than $c$ must be extremely close to $c$~\cite{Jacobson2008}.

The dispersion relation for a scalar field with Lagrangian density \eqref{lagrangian}
is $\omega^2 = v^2 k^2$ when expressed in terms of 
components of the wave 4-covector $k_a$ in the aether frame.
More precisely, the wave phase is $k_\mu x^\mu = \o t + k_ix^i$,
and $k^2 = \sum_i k_ik_i$. For real spatial wave vectors $k_i$,
the stability requirement that the frequency be real amounts to the condition
$v^2>0$. This condition guarantees that any solution that is a superposition
of plane waves on a constant $t$ surface is stable. 

One of the further stability criteria of Ref.~\cite{Carroll2009}  
is the demand that the frequency be real also for plane waves
on any Lorentz-boosted constant time surface. To determine
what that implies, we may reexpress the dispersion relation
in terms of the components of the wave 4-vector in the boosted frame
as follows.

The metric $\eta_{ab}$ can be used to define a set of
frames, related in the usual way by Lorentz transformations.
With respect to such a frame moving with velocity 
$\beta$, the new time and space coordinates are given by 
\begin{eqnarray}
t' &= & \gamma(t - \beta x_\|), \\
x'_\| &= & \gamma (x_\| - \beta t), \\
x'_\perp &= & x_\perp,
\end{eqnarray}
where ${}_\|$ and ${}_\perp$ 
refer to the components parallel and
perpendicular to the boost direction,
and we use units with the metric speed of light
equal to unity, $c=1$. 
The frame velocity will be taken to be positive, and is assumed
to be less than the speed of light, $0 \leq \beta < 1$.

The covariant (as opposed to contravariant)
frequency and wave 4-vector components
in the boosted frame are given by
\begin{eqnarray}
\omega' &= & \gamma(\omega + \beta k_\|), \\
k'_\| &= & \gamma (k_\| + \beta \omega), \\
k'_\perp &= & k_\perp.
\end{eqnarray}
The dispersion relation in terms of these boosted 
components takes the form
\beq \label{dispersion}
\begin{split}
(1 - v^2 \beta^2) \omega'^2 + 2 \beta k'_\| (1 - v^2) \omega' \qquad & \\
{} + (\beta^2 - v^2) k'_\|{}^2 - v^2 (1 - \beta^2)k'_\perp{}^2 &= 0,
\end{split}
\eeq
where we have multiplied by a factor $(1 - \beta^2) = \gamma^{-2}$ for convenience.
This is a quadratic equation for $\omega'$,
so the roots are real for real $k'$ if and only if the discriminant is positive,
\beq
v^2 ( 1 - \beta^2)^2 k'_\|{}^2 + v^2  (1 - v^2 \beta^2) (1 - \beta^2) k'_\perp{}^2 \geq 0.
\eeq
Since $\beta < 1$, this can be negative
only if the term $(1 - v^2 \beta^2)$ is 
negative, which occurs only if $v > 1$ and $\beta > 1/v$.
We note that this is just the condition for the constant $t'$ surfaces
to be timelike with respect to $g_{(v)}{}_{ab}$ \eqref{gv}.
The frequency $\omega'$ then has a nonzero imaginary part
when $k'_\perp/k'_\|$ is sufficiently large.

Thus for $v>1$ and $\beta > 1/v$
there exist solutions with real wave vectors and complex frequencies in the boosted frame.
Such modes grow exponentially in the time coordinate $t'$ of that frame.
Whether or not this indicates an instability comes down to the question
whether or not these solutions are part of the physical phase space of the theory.

As pointed out in \Cc, the wave vector $k_\|=\g(k'_\| -\b \o')$ in the rest
frame of the aether will be complex for such modes, so on a constant 
$t$ surface the solution will blow up exponentially at spatial infinity.
These solutions therefore do not satisfy the usual boundary conditions that define
the phase space of the theory on the constant $t$ slices. 
A consistent theory
can be defined by adopting a regular boundary condition on the 
constant $t$ slices, excluding these solutions. 
This is what is ordinarily done in a Lorentz invariant theory. 
For example, one could
require that the solutions have compact support, or that they be 
Fourier transformable on those slices. 
Moreover, since 
these boundary conditions are preserved by $t$ evolution, the theory so
defined preserves the time translation symmetry of the background.
Also, one would define the same phase space imposing these
boundary conditions on any other surface that is spacelike with respect to 
$g_{(v)}{}_{ab}$.

One may ask whether a consistent theory could instead be defined by adopting a regular boundary condition on the constant $t'$ slices. 
If so, this would raise the question of which is the correct phase space.
But it appears that this can not be done in a natural way.
For certain regular initial data on a given constant $t'$ surface the corresponding solution grows exponentially with $t'$.
In any other frame this solution will contain complex wave vectors and 
will therefore diverge asymptotically on the constant time slices of that frame.
This means that, unlike the case for surfaces which are spacelike with respect to $g_{(v)}{}_{ab}$, the phase space defined by regular data on constant $t'$ surfaces is different for every value of 
$\beta$ greater than $1/v$.

Moreover, for $\beta$ greater than 
$1/v$, the phase space obtained by requiring regular initial data on a fixed 
$t'$ slice depends not only on the particular value of $\beta$, but on the particular choice of slice.
For example, suppose the initial data on a particular surface $t'=t'_0$ possesses a well-defined Fourier transform.
In Fourier space, the wave equation then reduces to an infinite number of uncoupled ordinary differential equations that may be solved to obtain the Fourier transform of the solution on a different slice $t' = t'_1$.
For modes with sufficiently large $k_\perp$, the solutions to the differential equation grow exponentially with $k_\perp$.
The solution at $t'_1$ in Fourier space therefore 
does not in general have a convergent inverse Fourier transform.
Such initial data on the $t'_0$ surface do not correspond to 
any choice of initial data on the $t'_1$ surface.
This means that the phase space defined in this way depends 
not only on the choice of time coordinate $t'$ but also 
on the arbitrary value $t'_0$ of that coordinate, 
breaking the time translation symmetry of the theory.

Another reason to reject a ``$t'$-phase space" formulation is that allowing for arbitrary initial data at $t'=t'_0$ is unjustified in the context of a causal theory in which the 
$\vphi$ field interacts with other degrees of freedom. 
A simple way to see the problem is to allow for an external source term in the field equation for $\vphi$.
One can then ask whether the source could generate data at $t'_0$ that would lead to an exponentially growing solution.
The answer is no unless (perhaps) if the source is turned on in the infinite past.
As explained above, any such solution will blow up exponentially at spatial infinity on {\it all} constant $t$ surfaces.
If the source is turned on at a finite time, its effects cannot propagate any faster than $v$ in the aether frame, and so the solution can {\it not} blow up at spatial infinity at any finite time. 

The preceding argument depends on a choice of boundary condition
for the solution generated by the source, which is equivalent to the  
choice of Green's function for the wave equation. We implicitly adopted 
the retarded Green's function, which vanishes for $t<0$.
One might ask whether the argument would continue to hold using a $t'$-retarded Green's function that would vanish for $t'<0$.
It appears, however, that no such Green's function exists
for $\beta>1/v$. This can be seen as follows.
A standard method for constructing Green's functions is via the Fourier transform
\begin{equation} \label{green1}
G(t',x') \propto \int d^3k' d \omega' \frac{e^{i k' \cdot x'} e^{-i \omega' t'}}{(\omega' - \omega'_-)(\omega' - \omega'_+)}
\end{equation}
where $\omega'_{\pm}$ are the roots of the dispersion relation \eqref{dispersion}.
This integral can be performed along any contour that begins at $-\infty$ and ends at $+\infty$ along the real axis.
If the integral converges, then the wave operator acting on $G(t',x')$ can be moved inside the integral, canceling the denominator.
The remaining integrand has no poles, so the contour can be freely deformed to lie along the real axis, yielding a representation of the Dirac delta function.
To obtain the retarded Green's function, the $\omega'$ integral is performed along a contour that passes above all the poles, so that for $t'<0$ the integral vanishes.
For $t'>0$, both poles are enclosed by the contour that
can be closed in the lower half plane,
and the $\omega'$ integral yields
\begin{equation} \label{green2}
G(t',x') \propto \int d^3 k' e^{i k' \cdot x'} \frac{e^{-i \omega'_+ t} - e^{-i \omega'_- t} }{\omega'_+ - \omega'_-}.
\end{equation}
For large $k'_\perp$, the roots behave as $\omega'_\pm \sim \pm i k'_\perp$, so that the integrand in \eqref{green2} grows exponentially with $k'_\perp$ and the integral does not converge.
A $t'$-retarded Green's function therefore cannot be found by this
standard method, which strongly suggests that such a 
Green's function does not exist\footnote{Note that the exponential
instability alone does not account for the absence of a retarded
Green's function. For example, in the case of a 
tachyonic scalar field with a negative $m^2$,
the instability occurs only at low $k$, 
so the convergence of the Green's function is not spoiled,
and a retarded Green's function exists.}.

A second sign of possible instability discussed in \C 
is that the Hamiltonians generating $t'$ translations
can be unbounded below whenever $v \neq 1$.
In particular the Hamiltonian of linear perturbations
is unbounded below precisely when $\b > v$ or $\b > 1/v$
(the condition in \C was expressed in terms 
of the coupling constants of 
the theory rather than the mode speeds,
nevertheless the two conditions are equivalent).
This can be understood in a simple way as follows.

Let $\xi^a$ denote the $t'$ translation 4-vector. 
The Hamiltonian generating $\xi^a$ translations can be written
as an integral over an initial data surface $\Sigma$,
\begin{equation}
H_\xi = \int_\Sigma T^b{}_a \xi^a n_b\, d^3\Sigma,
\end{equation}
where $T^b{}_a$ is the canonical energy-momentum tensor
\begin{equation}
\sqrt{-\eta}\, T^b{}_a 
= \frac{\partial {\cal L}}{\partial (\partial_b \vphi)} \partial_a \vphi - {\cal L} \delta^b_a,
\end{equation}
$n_b$ is the unit normal covector, and $d^3\Sigma$ is the surface
volume element, both normalized with respect to $\eta_{ab}$. 
Positivity of $H_\xi$ is ensured when $T^b{}_a \xi^a n_b$ 
is positive. In the case $v<1$ the $\vphi$ field is ``subluminal" relative to 
$\eta_{ab}$, so for $\b>v$ the vector $\xi^a$ can be timelike 
relative to $\eta_{ab}$ but  {\it spacelike}  relative to the 
effective metric $g_{(v)}{}_{ab}$ for the $\vphi$ field. In this case
$H_\xi$ is in effect a component of the momentum,
not the energy of $\vphi$, which is clearly not bounded below.
In the case $v>1$, for $\b>1/v$ 
the $t'$ translation vector $\xi^a$ remains timelike with respect to $g_{(v)}{}_{ab}$, but 
the constant $t'$ surface becomes timelike with respect to $g_{(v)}{}_{ab}$.
In this case $H_\xi$ is the flux of energy through a timelike surface, and is no longer expected to be bounded below.
Moreover, if the surface $\Sigma$ is timelike with respect to $g_{(v)}{}_{ab}$
there is no guarantee that $H_\xi$ is conserved under $t'$ translation,
because the current $T^b{}_a \xi^a$ can flow out through the boundaries.

In conclusion, while we take no issue with the computations of \Cc,
the inference of instabilities in Einstein-aether 
theory when the mode velocities differ from $c$ is unwarranted.
A proper identification of the phase space of the theory eliminates
the exponentially growing solutions.
The Hamiltonians that were found to be unbounded below
actually correspond either to momenta or
to energy fluxes across timelike surfaces.
The Hamiltonian generating time translations in the 
aether frame is bounded below and plays the 
usual role of the energy in governing stability.
It is therefore sufficient for stability to impose the conditions
of real frequencies and positive energy in the aether frame.
The opposite conclusion was
reached in \C  by considering the Lorentz symmetry of the background
metric to be a physical symmetry of the phase space of linear perturbations. 
Since the background aether breaks this symmetry, that viewpoint is untenable. 

\section*{Acknowledgments}
We acknowledge helpful correspondence with S.~Carroll, T.~Dulaney, M.~Gresham, H.~Tam, C.~Eling and D.~Mattingly, 
This research was supported in part by the Foundational Questions Institute 
(FQXi Grant No. RFP20816), by NSF Grants 
No. PHY-0601800 and No. PHY-0903572, and by an NSERC PGS-D to WD.

\bibliographystyle{utphys}
\bibliography{aether-stability}

\end{document}